\documentclass[pre,showpacs,aps,twocolumn]{revtex4}
\usepackage{amsmath}
\usepackage{epsfig}
\begin{document}

\bigskip
\title{On the critical behavior of two-dimensional liquid crystals}

\author
{
Ana Isabel Fari\~nas-S\'anchez$^{\ddagger}$, 
Robert Botet$^{\star}$, 
Bertrand Berche$^{\star\star}$, 
and
Ricardo Paredes$^{\dagger}$
 }
 
\affiliation{
$^{\ddagger}$ Departamento de  F{\'\i}sica, Universidad  Sim\'on Bol{\'\i}var, 
Apartado Postal 89000, Caracas 1080-A, Venezuela\\
$^{\star}$ Laboratoire de Physique des Solides B\^{a}t.510, CNRS UMR8502 - Universit\'{e} 
Paris-Sud,
F-91405 Orsay, France  \\
$^{\star\star}$ Laboratoire de Physique des Mat\'eriaux, UMR CNRS 7556, Universit\'e 
Henri Poincar\'e Nancy 1, B. P. 239, F-54506 Vandoeuvre les Nancy Cedex, France.\\
$^{\dagger}$ Instituto Venezolano de Investigaciones Cient\'{\i}ficas, Centro de F\'{\i}sica, 
Laboratorio de Fisica Estad{\'\i}stica Apdo.  20632, Caracas 1020A, Venezuela}
 
\date{\today}
 
\begin{abstract}
The Lebwohl-Lasher (LL) model is the traditional model used to describe the 
nematic-isotropic transition 
of real liquid crystals. In this paper, we develop a numerical 
study of the temperature behaviour and of finite-size scaling of 
the two-dimensional (2D) LL-model. 
We discuss two possible scenarios. 
In the first one, the 2D LL-model presents a phase transition similar to the 
topological transition appearing in the 2D XY-model.
In the second one, the 2D LL-model does not exhibit any critical transition, 
but its low temperature behaviour is rather characterized by 
a crossover from a disordered phase to an ordered phase 
at zero temperature. 
We realize and discuss various comparisons with the 2D XY-model and the 2D 
Heisenberg model. 
Adding to previous studies of finite-size scaling behaviour of the order 
parameter and conformal mapping of order parameter profile, we analyze the
critical scaling of the probability distribution 
function, hyperscaling relations and
stiffness order parameter and  conclude that the 
second scenario (no critical transition) 
is the most plausible.

\end{abstract}



\maketitle 
\section{Introduction}

At high temperatures, the natural state of a liquid crystal is the 
isotropic phase,
because all the molecule orientations are equally probable. 
At low temperatures, an order along a preferred direction may appear, 
it is the nematic phase.
When such a material is cooled at even lower temperatures, other types
of ordered phases are likely to appear, the description of which requires  
realistic potentials (see e.g. reference \cite{Zewdie00}).
The transition between the isotropic and the nematic phases generally 
occurs at a critical temperature.
Various experiments on three-dimensional liquid crystals, exhibited a weak first order 
transition \cite{ZhangMouritsenZuckermann92}.
However, the corresponding two-dimensional (2D) problem is far from begin that clear. 
Experimentally, only 2D liquid crystals composed of long rigid interacting 
rods have been studied. 

First, the 2D problem must be defined through the symmetries: there are generally
a continuous global symmetry $O(n)$, and a local $Z_2$ symmetry. For example, the two most-studied models
deal with $O(2)$ (the XY-model), or $O(3)$ (the Heisenberg model) symmetries. The Hamiltonian reads in both cases as:
\begin{equation}
{\cal H}=-J \sum_{\langle i,j \rangle } {{\bf S}_{i} \cdot {\bf S}_{j}} 
\label{eq:intro1},
\end{equation}
with ${\bf S}_{i}$ a  $n$-components spin, of unit length, at the lattice 
site $i$. 
The sum runs over all pairs of neighbouring spins of the 2D square lattice and 
$J > 0$ is the spin-spin coupling constant.

According to the Mermin-Wagner-Hohenberg 
theorem \cite{mw,h,mermin1,mermin,coleman}, 
the order 
parameter fluctuations
prevent from 
any spontaneous breakdown of continuous symmetries for  2D systems 
with short-range interactions. 
However, this theorem cannot conclude in the case of transitions to a phase
with no long range order. 
An example of such a 2D critical behaviour, which is not a continuous-symmetry 
breaking, is the 2D XY-model
($O(2)$, Abelian symmetry). 
This model presents a  topological transition 
characterized by a critical low temperature phase 
as a plasma of vortices pairs, 
at temperatures under the Berezinskii-Kosterlitz-Thouless transition 
temperature, 
$T_{BKT}$ \cite{bere,kt}. At much lower temperatures, the system state 
is dominated by spin waves.   The high temperature phase occurs when the pairs 
dissociate.

On the other hand, the 2D Heisenberg model ($O(3)$, non-Abelian symmetry) 
has no transition at any finite temperature 
(asymptotic freedom) \cite{polya,brez,amitbook,izyu}. 
The longitudinal 
modes of the $O(3)$ model are frozen 
for the low temperatures, and only the transverse modes are activated, 
leading essentially to two Gaussian modes. The deep difference
between the two models at finite temperatures, is the stability of the 
vortices. Indeed, the topological 
transition of the 2D XY-model is the result of the stability of the 
topological defects, 
while the ``third spin dimension"  in the $O(3)$ model, makes the vortices 
unstable at any finite temperature.

However, the above picture has been questioned recently by the
 numerical evidences  
for a possible transition 
in the 2D Heisenberg 
model \cite{Niedemayer,PatrasciouSeiler,Patrasciou,Aguado}. 
Moreover it has been shown that 
a  quasi long range order (QLRO) phase \cite{tesisAna,bb} 
might appear at very low temperatures in finite systems. 
Indeed in these numerical works, the thermodynamic limit remains questionable
and the analytic approach of \cite{bb} relies explicitly
on the finite size of the lattice.

A model for a regular 2D liquid crystal system 
was proposed by Lebwohl and Lasher (LL) \cite{ll}. Based on 
a lattice version of the mean field theory of Maier and 
Saupe \cite{MaierSaupe59}, the molecules are represented by 
$n-$component unit vectors and
the ``spin-spin" interactions
are given by the second Legendre polynomial, $P_2$ 
(keeping the local $Z_2$ symmetry). 
The popularity of the LL-model comes from its ability to reproduce the 
weak first order phase transition observed 
experimentally between the isotropic and the nematic 
phases \cite{ZhangMouritsenZuckermann92} in the three-dimensional space. 
Then, the question of possible phase transitions in this model is  
attractive and has been addressed in numerous 
studies. \cite{tesisAna,kuzu,blote,caracciolo,FPB,PFB}

In the past twenty years, several numerical evidences pointed out a 
possible topological transition 
for the 2D nematic-isotropic phase transition in liquid crystals. 
The situation is clear in the case of planar rotator models (see e.g. 
references \cite{FPB2,BP}), but still controversial for three-component models.
In reference \cite{FPB2}, we 
have studied the nematic-isotropic transition 
of 2D liquid crystals using a $O(2)$ vector model 
characterized by nonlinear nearest-neighbour spin interactions 
governed by the fourth Legendre polynomial $P_4$. 
The system has been studied through standard finite-size scaling 
and conformal rescaling of the density profiles. We also estimated the Binder 
cumulant \cite{Binder} as a function of the temperature for 
different values of system size $L$. 
This cumulant has been proved to be universal at a critical temperature. 
Evidences for a topological transition at a finite temperature 
have been underlined.

In the case of the three-component model, in
a remarkable work Kunz and Zumbach \cite{kuzu}  
concluded in favour of 
such a critical topological transition.  
Although they performed a careful and sizable study, 
they were unable to decide conclusively between essential 
singularities or standard power 
laws for the correlation length and the susceptibility, 
when approaching the critical temperature 
from the high temperature phase. However, 
they developed a qualitative picture 
(pairing of topological defects which carry most part of the system energy, 
sharp increase of the defects density and seeming 
discontinuity of the rotational-rigidity modulus, 
finite cusp of the specific heat, 
proliferation of unbounded defects at high temperature) and
their conclusion inclined clearly to the essential singularity,
for a temperature $T_{BKT}$.
More recently, we studied the 2D LL-model \cite{tesisAna,FPB,PFB}, 
and found signals in favour of a QLRO phase at $T\le T_{BKT}$, with  
a magnetization that decays with the system size as a 
power law with the critical exponent 
$\eta(T)/2$ \cite{PFB}. We obtained good agreement with the 
$\eta(T)$ exponent obtained using the powerful 
technique of the conformal transformation \cite{FPB}. 
In \cite{tesisAna}, the value of
$\eta(T)$ was estimated using directly the susceptibility, and the 
agreement appeared to be reasonable. 
Possible discrepancy could come from the finite size of the systems. This QLRO phase could be 
 similar to the low temperature phase of the 2D XY-model. 
Mondal and Roy \cite{mr}, using standard finite size scaling method and 
Monte Carlo simulations for the 2D LL-model, 
gave arguments for a continuous phase transition. The critical temperature 
was estimated to be $T_c = 0.548 \pm 0.02$, 
a little higher than $0.513$ calculated in \cite{tesisAna,kuzu,FPB,PFB}. 
A still more recent study reported evidences in favour of a topological 
transition \cite{DuttaRoy04}. 
Then, the transition might be driven by stable topologically point 
defects known as $\frac 12$-disclination points.

Despite 20 years of numerical results claiming for a topological transition 
for the 2D LL-model,
a recent work called seriously this scenario into question. Indeed, precise 
estimates of 
the Binder cumulant for various system sizes, have been unable to show any 
crossing point at a definite temperature \cite{RAR}. 
This result contradicts completely the transition scenario at a critical 
point. 
In the same work \cite{RAR}, the possibility of a QLRO phase 
at $T=0.4$ (a value well below the ``transition" temperature 
estimated previously for this model) was questioned. Two strong 
evidences were given to conclude that the 2D Lebwohl-Lasher 
model does not show any quasi-long-range ordered phase.
The evidences were the violation of the hyperscaling relation 
between the apparent exponents of the magnetization and the 
susceptibility, and the failure of the 
first-scaling collapse \cite{bp} of the probability distribution function of the order parameter.

The main goal of the present work is to present a complementary 
investigation of the 2D LL-model.  
We will compare with the same tools this model with the 2D XY- and 
Heisenberg models, for which the 
critical behaviours are not questionable. 
Since the 2D LL-model seems to share some common properties with both the XY- 
and Heisenberg models, it appears helpful to provide the  comparisons.
In section  \ref{orderparameter} 
we will analyze the system with and without an applied magnetic field
using the finite size scaling method (FSS). The results which follow from
conformal 
transformation method (CT) will also be reminded \cite{FPB}. 
We shall recover the apparent evidences for a sort of topological 
transition similar to the 
Berezinskii-Kosterlitz-Thouless transition. However, we will see in the 
section  \ref{sec:PDFhyper}, with new results on
the shape of the tail of the probability distribution function for the 
magnetization, 
check of the hyperscaling relation, and study of the stiffness, how 
improbable is the reality of
a critical topological transition, confirming the results of \cite{RAR}.

\section{Definitions of the models}

We will consider hereafter the 2D XY-model \cite{bere}, 
 the 2D Heisenberg model \cite{polya}, and the 2D Lebwohl-Lasher 
model \cite{ll}.
The simulations will be performed using Monte Carlo (MC) 
Wolff cluster algorithm \cite{wolff}
appropriately adapted to the model under consideration \cite{kuzu},
in order to avoid most of the critical slowing down behaviour close to 
transition, if any.
All simulations are realized on a square lattice of size $L \times L$ with 
periodic boundary conditions. 
The system sizes and number of MC iterations will be specified when needed
in the figure captions. The constants $J$ and $k_B$ are fixed to unity.

\begin{itemize}
\item
For the 2D XY-model, the $N=L^2$ classical spins are confined in the lattice 
plane $x, y$. 
The spin ${\bf S}_i$ is parameterized by: 
$(S_i^x = \cos \theta_i, S_i^y=\sin \theta_i )$, with
$\theta_i$ the angle between the direction of the spin ${\bf S}_i$ and 
the $x$-axis. 
According to (\ref{eq:intro1}),  the system state is governed by the 
Hamiltonian 
${\cal H}=-J\sum_{\langle i,j \rangle} \cos (\theta_i -  \theta_j )$, 
with the ferromagnetic coupling constant  $J>0$,
and the sum running over all the nearest-neighbour pairs of spins. 
The critical features are eventually characterized by the 
power-law behaviour of the 
magnetization per site: $m\equiv\frac{1}{N}\sqrt{(\sum_i {\bf S}_i)^2}$, 
which is a non-negative real number. 
Thermal fluctuations of the order parameter give access to the 
susceptibility $\chi=L^d/k_BT(\langle m^2\rangle-\langle m\rangle^2)$.

\item
For the Heisenberg model, the $N=L^2$ classical spins can point in any direction
of the three dimensional space $x, y, z$,
$(S_i^x = \sin \theta_i\cos\varphi_i, S_i^y=\sin \theta_i\sin\varphi_i, 
S_i^z=\cos \theta_i )$, with the standard definitions for the local angles.
Similar to the above XY-model, the 
Hamiltonian has the form  
${\cal H}=-J\sum_{\langle i,j \rangle} {\bf S}_i \cdot {\bf S}_j $, 
but with three-component spins and the local order parameter is defined
by the appropriate magnetization per site.

\item
For the Lebwohl-Lasher model, the $N=L^2$ classical spins can also 
point in any 
of the three dimensional space directions $x, y, z$.
The Hamiltonian of the system, 
\begin{equation}
{\cal H}=-J\sum_{\langle i,j \rangle} P_{2}( {\bf S}_i \cdot {\bf S}_j),
\end{equation}
is an obvious generalization of the Heisenberg 
interaction which guarantees the local $Z_2$ symmetry of liquid crystals.
$P_2$ is the second Legendre polynomial, that is: $P_2(x) = (3 x^2  -1)/2$. 
One defines the local order parameter (also called nematization) 
as: $m \equiv\frac 1N\sum_i P_2(\cos\alpha_i)$, 
where $\alpha_i$ 
is the angle between the direction of the spin ${\bf S}_i$ and the
direction of symmetry breaking.

\end{itemize}

\section{Qualitative  Analysis of the Order Parameter and the Susceptibility}\label{orderparameter}

In the following we study the behaviour of the order parameter
(magnetization or nematization) 
$m$ and the magnetic susceptibility (or response function) $\chi$, 
as a function of the temperature for various system sizes $L$. 

\subsection{The 2D XY-model}
\begin{figure}[!htb]
\centerline{\includegraphics[width=0.45\textwidth]{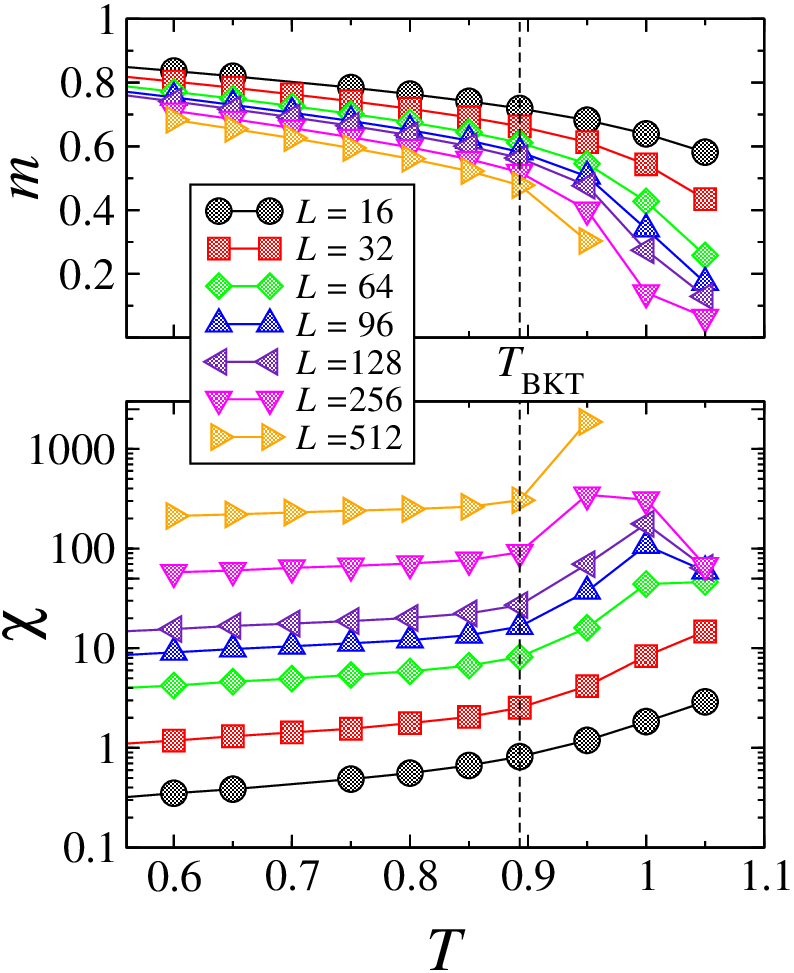}}
\caption{Temperature dependence of thermodynamic quantities for the 2D XY-model. 
Top: magnetization (m). Bottom: susceptibility ($\chi$). 
(Monte Carlo simulations with $10^6$ MCS/spin, after removing the first $10^6$ 
MCS to achieve proper thermalization).}
\label{fig:fssxy}
\end{figure}

In figure  (\ref{fig:fssxy}), 
the magnetization $m$ and the susceptibility $\chi$ (top and bottom) 
for the 2D XY-model are plotted versus the temperature.
Both plots have the same temperature range, wide enough to cover the domain of temperatures 
where important thermodynamic changes are expected to occur (essentially: the BKT transition, which should appear
at the temperature $T_{BKT}\simeq 0.893$).

\begin{figure}[!htb]
\centerline{\includegraphics[width=0.45\textwidth]{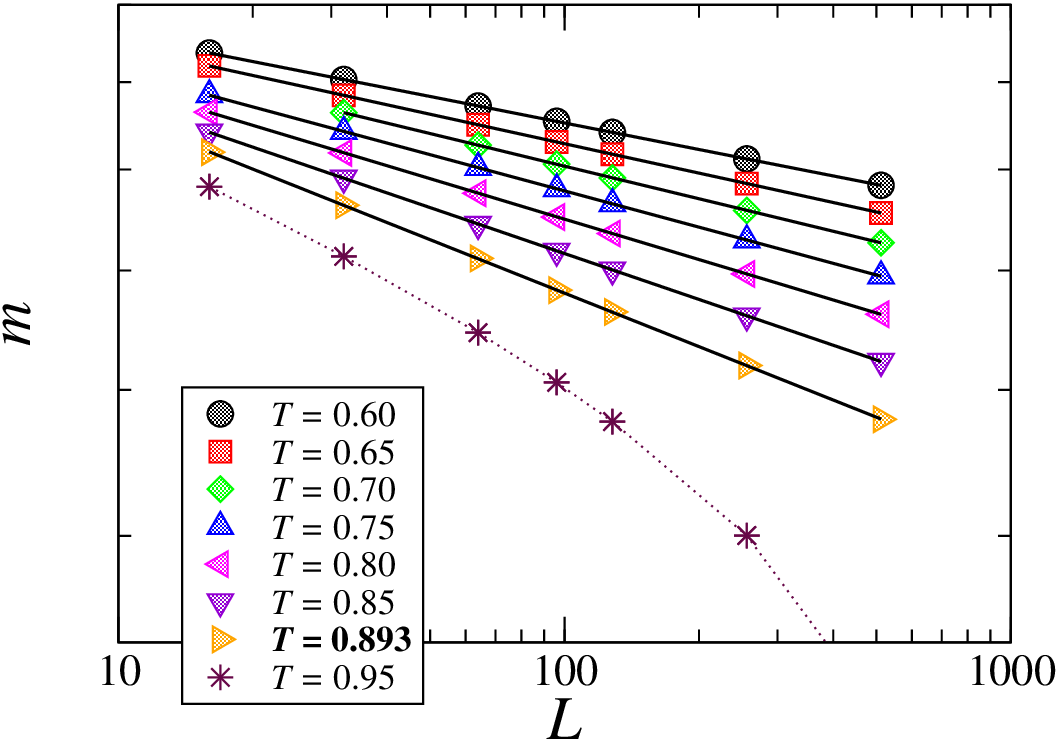}}
\caption{Behaviour of the magnetization versus the system size 
at different temperatures for the 2D XY-model. Power-law behaviours appear
in the whole range $T\le T_{BKT}$.
(Monte Carlo simulations with $10^6$ MCS/spin after removing the first $10^6$ 
MCS to achieve proper thermalization).}
\label{fig:mvslxy}
\end{figure}

We can observe in  figure  (\ref{fig:fssxy}) a progressive decrease of the 
magnetization with the system size,
at constant temperatures. 
Moreover, as become evident from figure  (\ref{fig:mvslxy}), 
we see that these decays 
follow power-laws with the system size 
in the whole range $T\le T_{BKT}$, according to the FSS picture in the critical
phase, $m\sim L^{-\frac 12\eta(T)}$, with an exponent $\eta(T)$
which depends on the value of the temperature (to be discussed in the next
section). 
When the value of the temperature is above $T_{BKT}$, 
the decay of the magnetization is faster than a power law.

\begin{figure}[!htb]
\centerline{\includegraphics[width=0.45\textwidth]{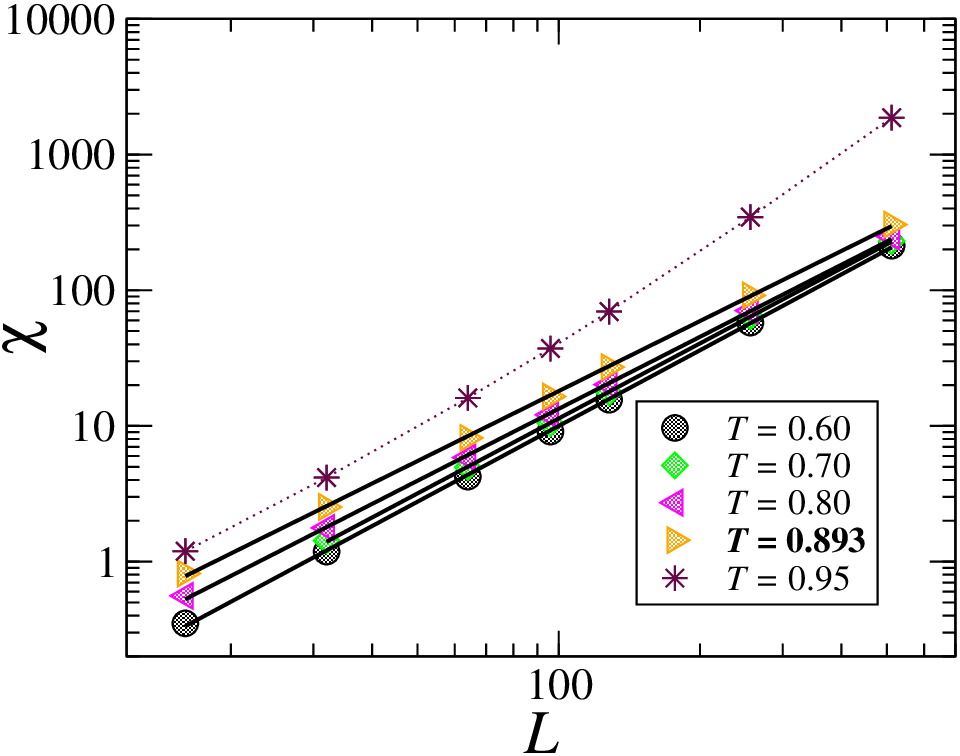}}
\caption{Behaviour of the magnetic susceptibility versus the system 
size at various temperatures for the 2D XY-model. (Monte Carlo simulations with 
$10^6$ MCS/spin after removing the first $10^6$ 
MCS to achieve proper thermalization).}
\label{fig:chivslxy}
\end{figure}

On the other hand, the magnetic susceptibility at
fixed temperature (figure  \ref{fig:fssxy}) 
exhibits regular increase with the
system size for all temperatures at and below $T_{BKT}$. It is in contrast 
with an ordinary second order phase transition 
in which the susceptibility diverges only at the critical temperature. 
In particular, the susceptibility exhibits power law behaviours 
with $L$ for any $T\le T_{BKT}$ (figure  \ref{fig:chivslxy}).
It is interesting to notice that a change in the  exponent of the power-law
of the susceptibility with the system size, appears at $T_{BKT}$. 

Also, we see in figure  (\ref{fig:fssxy}), that the susceptibility reaches its maximum value systematically above $T_{BKT}$. 
The temperature at which this maximum occurs tends to $T_{BKT}$ when $L\rightarrow \infty$. 

The power law behaviours of the magnetization and of
the magnetic susceptibility at low temperatures, can be interpreted as 
evidences of a QLRO phase for the 2D XY-model in the thermodynamic limit, 
in the temperature 
range $T < T_{BKT}$.

\subsection{The 2D Heisenberg model}
The 2D Heisenberg model is expected not to exhibit any phase 
transition \cite{polya,brez,amitbook,izyu}. 
The magnetic susceptibility strongly diverges 
when the  temperature vanishes \cite{polya}. 
However, some authors argued in favour of a 
transition \cite{Niedemayer,PatrasciouSeiler,Patrasciou,Aguado} 
and possibly an effective
quasi-long-range-ordered phase \cite{tesisAna,bb} at very low temperatures,
at least in large but finite systems. 

\begin{figure}[!htb]
\centerline{\includegraphics[width=0.45\textwidth]{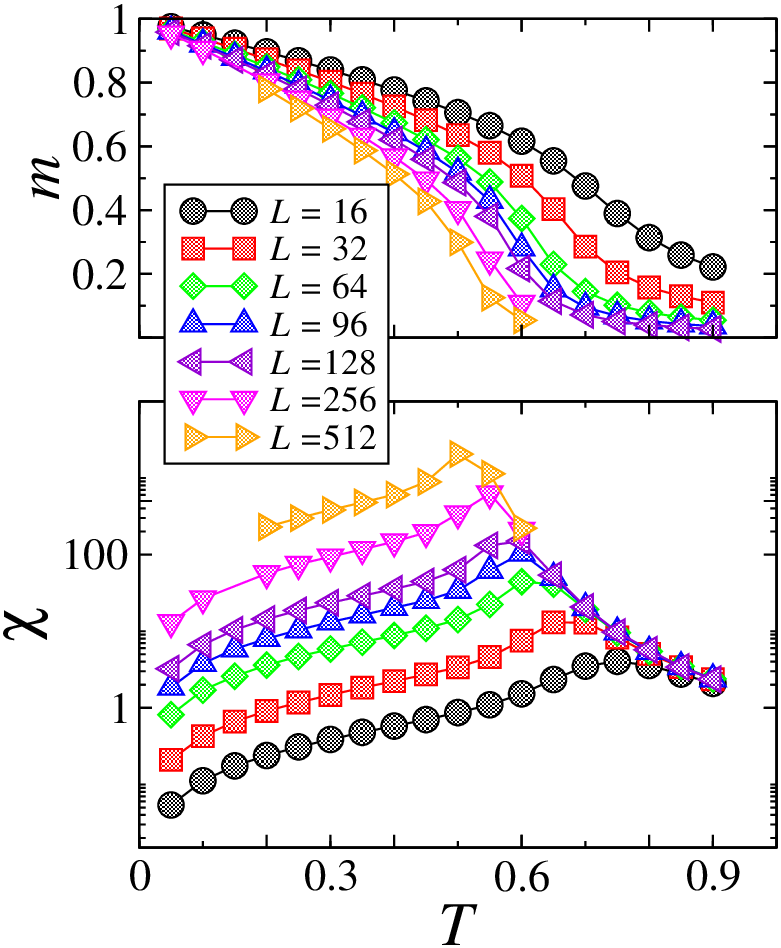}}
\caption{ Temperature dependence of thermodynamic quantities for the 2D Heisenberg model. 
Top: magnetization (m). Bottom: susceptibility ($\chi$). 
(Monte Carlo simulations with $10^6$ MCS/spin after removing the first $10^6$ 
MCS to achieve proper thermalization).}
\label{fig:fssheisenberg}
\end{figure}

In figure  (\ref{fig:fssheisenberg}), one can see the behaviours of the magnetization and of the magnetic susceptibility
with the system size and temperatures for the 2D Heisenberg model. We can notice the decrease of the values of the 
inflection point of $m$ versus $T$, when the system size increases. On the same way, the temperatures
where the susceptibility reaches the maximum, are displaced to lower and lower values when the system sizes 
are larger. As a matter of fact, if there is a positive critical temperature analogous to the 
$T_{BKT}$, its value should be quite small, at the thermodynamic limit.

\subsection{The 2D LL-model}
\begin{figure}[!htb]
\centerline{\includegraphics[width=0.45\textwidth]{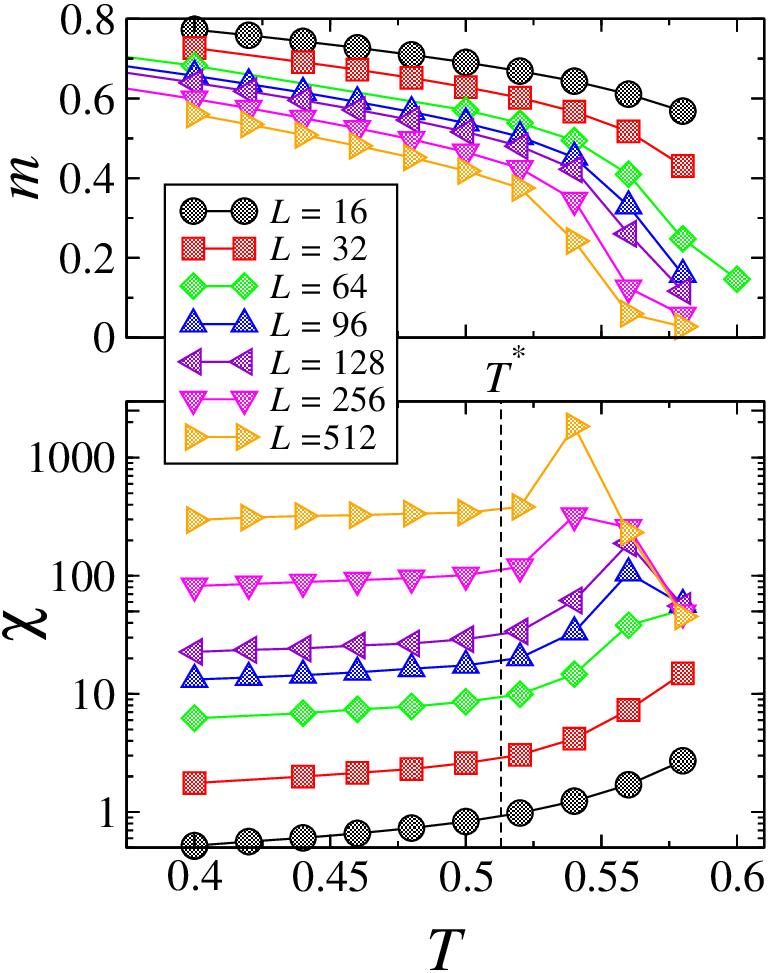}}
\caption{Temperature behaviour of thermodynamics quantities for the 2D Lebwohl-Lasher model. Top: magnetization (m). 
Bottom: magnetic susceptibility ($\chi$). 
(Monte Carlo simulation with $10^6$ equilibrium steps (measured as the number of flipped Wolff clusters) after removing the first $10^6$ MCS to achieve proper thermalization).}
\label{fig:fssll}
\end{figure}

\begin{figure}[!htb]
\centerline{\includegraphics[width=0.45\textwidth]{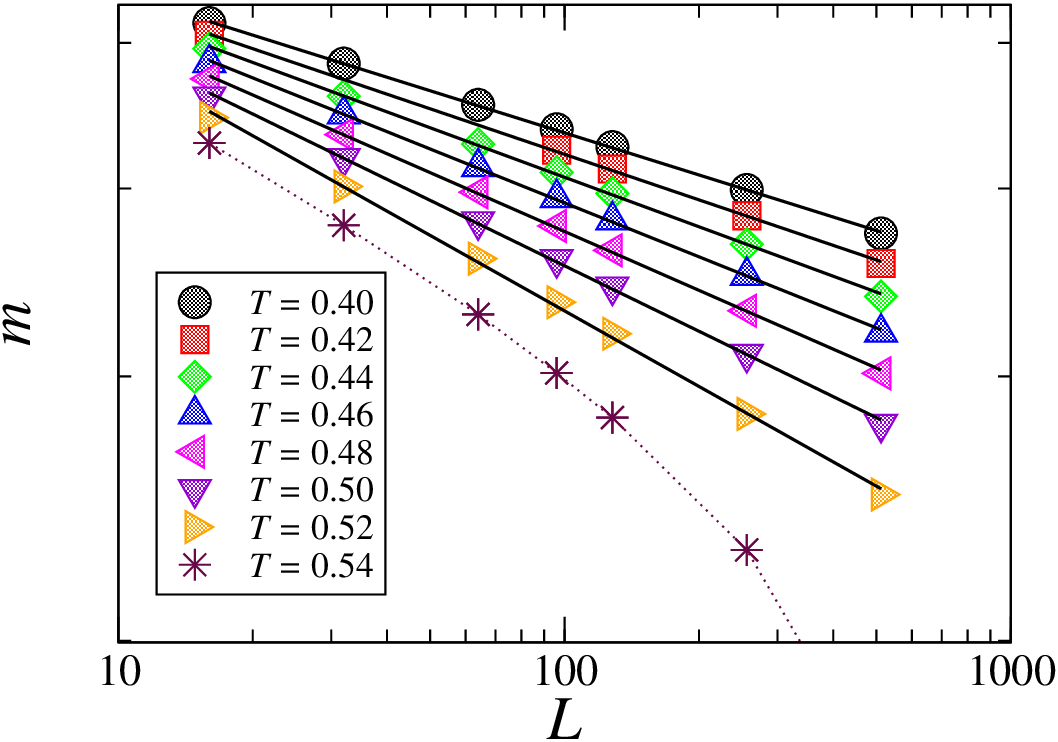}}
\caption{ Behaviour of the magnetization with the system size 
at various temperatures for the 2D LL-model. 
(Monte Carlo simulation with $10^6$ equilibrium steps (measured as the 
number of flipped Wolff clusters) after removing the first $10^6$ 
MCS to achieve proper thermalization).}
\label{fig:mvslll}
\end{figure}

\begin{figure}[!htb]
\centerline{\includegraphics[width=0.45\textwidth]{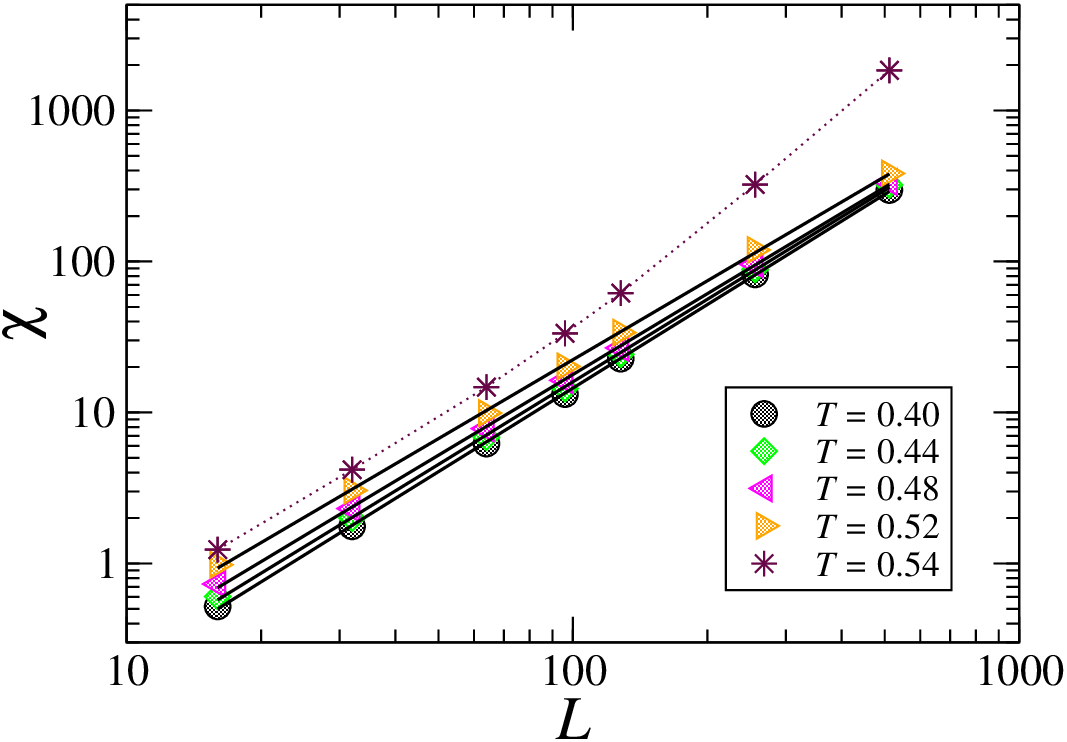}}
\caption{Behaviour of the magnetic susceptibility versus the system size 
at various temperatures for the 2D LL-model. 
(Monte Carlo simulation with $10^6$ equilibrium steps (measured as 
the number of flipped Wolff clusters) after removing the first $10^6$ 
MCS to achieve proper thermalization).}
\label{fig:chivslll}
\end{figure}

We analyze now the magnetization and the magnetic susceptibility of the 2D Lebwohl-Lasher model, 
using the same procedure as for the 2D XY-model. In figure  (\ref{fig:fssll}) the magnetization, $m$, 
and the magnetic susceptibility, $\chi$, are shown versus the temperature for various lattice sizes $L$. 
There is a complete qualitative agreement with the behaviour
 of the same quantities for the 2D XY-model (figure  \ref{fig:fssxy}). 
Indeed the ``critical" temperature is shifted to $T^\star=0.513$ \cite{kuzu,FPB}, but all the 
power-law behaviours are equally recovered (figures  \ref{fig:mvslll} 
and \ref{fig:chivslll}), 
and even the change, at $T^\star$, of the apparent value of the
exponent of the susceptibility versus the system size, or the asymptotic convergence to $T^\star$,
of the temperatures at which the susceptibility reaches its maximum are identically
noticed.

\subsection{Conclusion from the study of the $m$ and $\chi$ behaviours}
From this preliminary analysis, one could conclude  
that the 2D LL-model behaves similarly to the 2D XY-model. 
Namely, one can define a BKT temperature, $T^\star$, at which the system is 
critical, and a  QLRO phase
seem to occur at temperatures smaller than $T^\star$. 

We shall see below that this anticipated conclusion 
is refuted by more refined analysis.

\section{Calculations of the correlation function exponent $\eta$}

In principle, very precise information can be obtained from the estimates of the 
fundamental critical exponent $\eta$ for the pair correlation function. 
This exponent may depend on the effective temperature $T$ in the case of a continuous line of critical points. 

Several methods can  be used independently in the case of the BKT transition.
We shall use the following ones below:

\begin{itemize}
\item
Finite size scaling  behaviour of the magnetization. 

At the critical point of a second order phase transition, 
the magnetization $m$ behaves as a power law of the system size, namely: 
$m \sim L^{-\beta/\nu}$. This combination of critical 
exponents is related to the
exponent which describes the critical
decay of appropriate pair correlation function, $G(r)\sim r^{-(d-2+\eta)}$,
$2\beta/\nu =d-2+\eta $, thus in two dimensions $m\sim L^{-\frac 12\eta}$.
Here, $\eta$ is a function of $T$ and FSS is used to estimate 
the critical exponents at different temperatures (in the low temperature 
phase) from  
figure  (\ref{fig:mvslxy}) or figure  (\ref{fig:mvslll}) ;

\item
FSS behaviour of the magnetic susceptibility.

The magnetic susceptibility, $\chi$, scales as a power law 
involving  the critical exponents $\gamma$ and $\nu$, namely:
$\chi \sim L^{\gamma/\nu}$. Then, the hyperscaling 
relation\footnote{This scaling law is obtained using e.g. Rushbrooke equality
$\alpha+2\beta+\gamma=2$ and the hyperscaling relation in its standard form,
$\alpha=2-d\nu$ and the definition of the exponent $\eta$ through
$d-2+\eta=2\beta/\nu$. It also simply follows from the definition of $\chi$
in terms of the correlation function, $\chi_L=\int_{L^d}G(r)d^dr$.} 
\begin{equation}
\label{eq:hyper}
\eta = 2 - \gamma/\nu,
\end{equation}
 holds, which gives an alternative estimate of the exponent $\eta$.
Here too, FSS may be used to extract an appropriate exponent 
from the data of figure  (\ref{fig:chivslxy}) or figure  (\ref{fig:chivslll}); 

\item
Conformal Transformation (CT) method.

Another tool can be used to estimate the value of $\eta$ for very large 
systems: the conformal transformation method (see Cardy \cite{Cardy}).
It is indeed a powerful method which applies to analyze 
any conformally covariant density profile (or correlation function) from
the mapping of infinite or semi-infinite geometries onto
restricted geometries where simulations are actually 
performed \cite{berche03,ResStraley00}.
We thus consider a density 
profile $m(w)$ in a finite system with symmetry breaking fields along some 
edges in order to induce a non-vanishing local order parameter in 
the 2D bulk of the system. Here, $w$ stands for the location in the
finite system.
In the case of a square lattice  of size 
$L \times L$, with fixed boundary conditions along the four 
edges, using a simple mathematical 
transformation -- called 
the Schwarz-Christoffel transformation \cite{derridaburkhardt,chatelain} -- 
one expects a power law in terms of an \emph{effective} locally rescaled 
distance
\begin{eqnarray}
	m(w)& \sim&
 	[\kappa(w)]^{-\frac{1}{2} \eta}.
	\label{eq:squaremapping}
\end{eqnarray}

This method has been used extensively in \cite{berche03,EPL} 
to estimate the pair correlation exponent as a function of temperature for the 
2D XY-model;

\item
Systems in an applied magnetic field.

According to the scaling hypothesis, 
the magnetization of a general $d$-dimensional system at a critical temperature, must follow the scaling law:
\begin{equation}
m = H^{1/\delta}G(L^{-1}H^{-1/y_h}),\label{eq:universal}
\end{equation}
where  $H$ is the applied magnetic field. 
The magnetic exponent is given in terms of the magnetic field scaling dimension
$y_h$,  
$\delta=y_h/(d-y_h)$, and $G$ is a universal function \cite{golden}.
In the thermodynamic limit $L^{-1}H^{-1/y_h}\ll 1$, the system follows the singular behaviour $m\sim H^{1/\delta}$.
In the other limit $L^{-1}H^{-1/y_h}\gg 1$, the magnetization follows $m \sim L^{y_h-d}$. As we know 
that  $m \sim L^{-\eta/2}$ for the BKT transition, one deduces in this case the  hyperscaling relation under the form:
\begin{equation}
d-y_h=\eta/2.\label{eq:hyper2}
\end{equation}
which again can be used to estimate the value of $\eta$.

\end{itemize}

\subsection{The 2D XY-model}
The FSS methods are used from figure   (\ref{fig:mvslxy}) and figure   (\ref{fig:chivslxy}). The results obtained by the CT method
are from the reference  \cite{EPL}.
\begin{figure}[!htb]
\centerline{\includegraphics[width=0.45\textwidth]{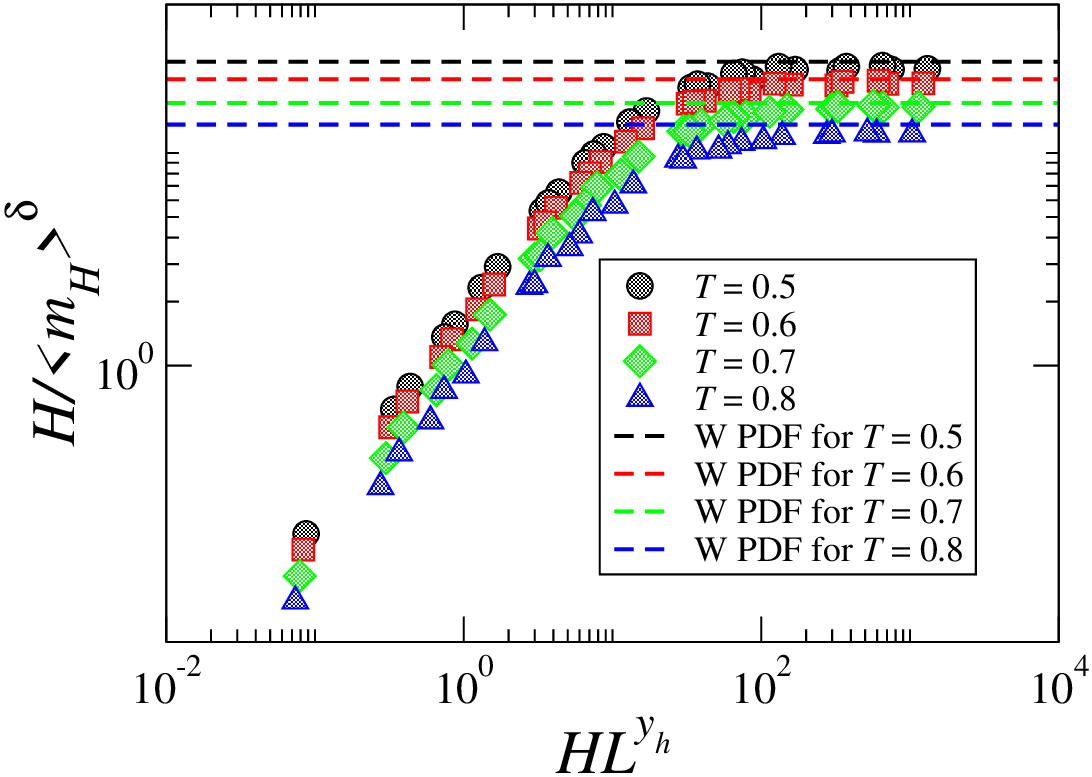}}
\caption{XY-model in a  magnetic field H. For each temperature the system sizes are: $L = 32$, $64$, $96$, $128$, 
and the magnetic field strengths are: $H = 0.1$, $0.01$, $0.05$, $0.001$, $0.005$, $0.0001$, $0.0005$.  
The bold straight lines represent the large $X=HL^{y_h}$ limit for a Weibull-like PDF (W) for each temperature. 
The number of Wolff steps for each $T$, $L$ and $H$ is $2\times 10^5$. }
\label{fig:fig4xy}
\end{figure}
We add here a few words about the 2D XY-model in a magnetic field. 
The exponent $\eta$ at the BKT transition is $1/4$, 
then $y_h=15/8$, and $\delta=15$ has the same value as for the $d=2$ Ising model.
Using equation (\ref{eq:universal}), an excellent collapse 
of $ Y = H/m^\delta$ versus  $X = HL^{y_h}$ at $T_{BKT} =0.893$ has been obtained in \cite{rprb}. The authors 
obtained in this work the additional result that  $Y$ tends to a constant value in the limit
$HL^{y_h}\rightarrow \infty$. 
The saturation of $Y$ for large magnetic fields
means a Weibull-like fieldless probability distribution function (PDF) \cite{bouk}
for the magnetization. We extended in the present work this result to other temperatures.

\begin{figure}[!htb]
\centerline{\includegraphics[width=0.45\textwidth]{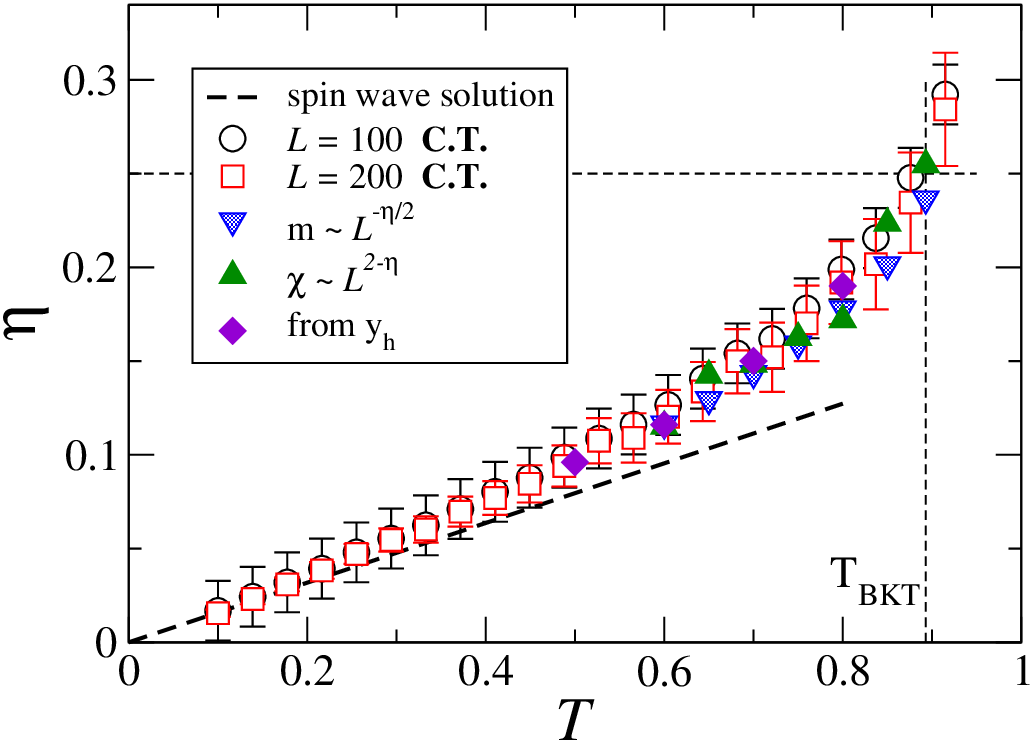}}
\caption{Dependence of the correlation function critical 
exponent with the temperature 
for the 2D XY-model.  
(Monte Carlo simulations with $10^6$ MCS/spin after removing the first $10^6$ 
MCS to achieve proper thermalization). 
The  conformal transformation (CT) values are from the reference \cite{EPL}.
The dashed line is the spin-wave result $\eta=k_BT/2\pi J$.}
\label{fig:etaxy}
\end{figure}

Indeed, for a BKT transition there is a line of critical points for any $T\le T_{BKT}$, and the
equation (\ref{eq:universal}) must be satisfied in all this range of temperatures.
Then, we performed numerical simulations of the 2D XY-model in an applied magnetic field \cite{wolffH} 
for temperatures below $T_{BKT}$. Figure  (\ref{fig:fig4xy}) shows 
the excellent collapse of the data, as well as the saturation of $Y$ for several magnetic field strengths 
and several system sizes, at all the temperatures $\le T_{BKT}$ investigated. 
The collapse of the data gives an estimate of  the critical exponent $\eta(T)$ 
using the hyperscaling relation (\ref{eq:hyper2}).

This result invalidates a claim by Bramwell {\sl et al.} \cite{bram1},
about the possible Gumbel-like distributions \cite{bouk} of the magnetization in the 2D XY-model at low temperatures. 
Indeed, the Gumbel PDF does not lead to the saturation of the function $Y$ in the limit $HL^{y_h}\rightarrow \infty$.

The figure  (\ref{fig:etaxy}) shows the values of the critical exponent $\eta$ for the 2D XY-model, 
calculated from FSS using the magnetization or the magnetic susceptibility, 
with the CT method (from \cite{EPL}), and using the collapse of the data for the system in a magnetic field. 
All the  methods lead to determinations of $\eta$ which 
are in excellent agreement with each other.

However, one can note that the behaviour obtained with CT is the most precise, because this method 
is able to reach the behaviour of the infinite-size systems, 
with system sizes computationally accessible, all shape effects being encoded
(in the continuum limit) in the conformal mapping (hence the name
of Finite Shape Scaling that we tried to introduce in Ref. \cite{FPB2}). 
The differences between the results obtained from $m$ and from $\chi$, are of the order of $5\%$. 
In the references \cite{RAR} and \cite{rprb}, the authors used more than $6 \times10^6$ independent realizations 
for the estimate of the exponent $\eta$. Realization of the hyperscaling relation (\ref{eq:hyper}) 
has been checked with a relative error smaller than $0.4\%$.

More than simple illustrations, these results allow us to compare the 
effeciency of the different methods for the proper definition 
of a quasi-long-range-order phase in systems with continuous symmetry. 

\subsection{The 2D Heisenberg model}
We compare now the values measured for $\eta$ for 
a system (the 2D Heisenberg model) 
which does not exhibit any transition, 
at least for temperatures above $T=0.1$ \cite{polya} (see 
figure  \ref{fig:fssheisenberg})). 
Then, we can expect the behaviours of $\eta$ to be quite different in the two 
cases.

\begin{figure}[!htb]
\centerline{\includegraphics[width=0.45\textwidth]{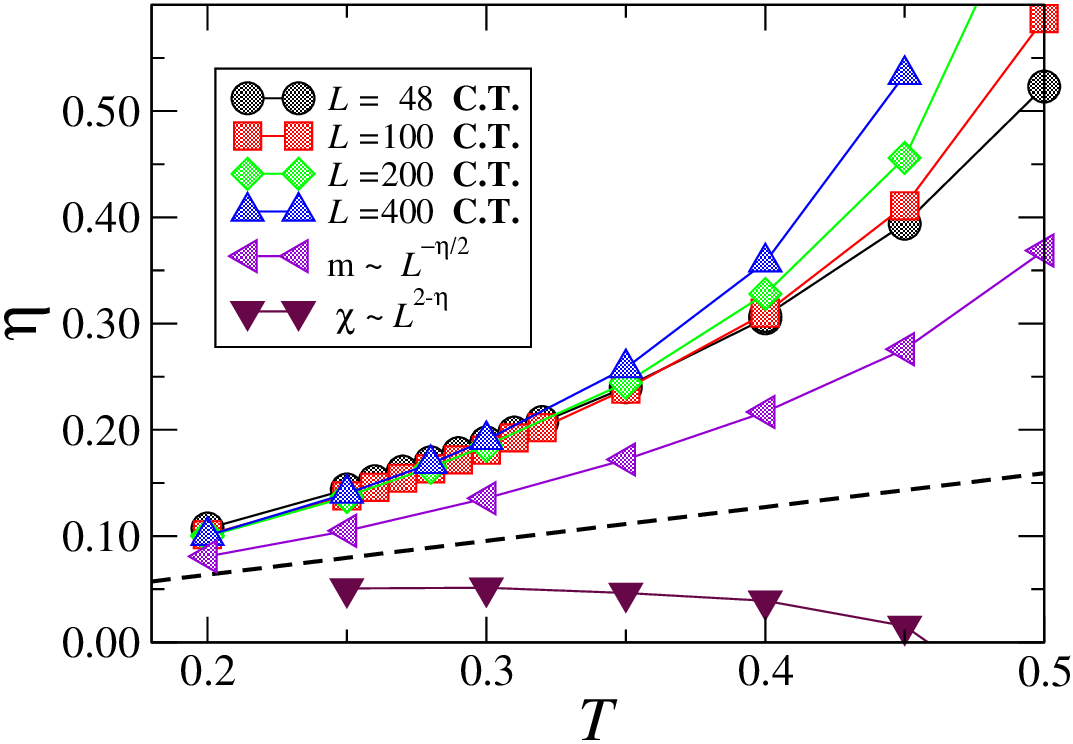}}
\caption{Dependence of the effective correlation function critical exponent 
with the temperature for the 2D Heisenberg model.
The dashed line is the spin-wave result $\eta=k_BT/\pi J$ \cite{bb}.}
\label{fig:etaheisenberg}
\end{figure}

The figure  (\ref{fig:etaheisenberg}) shows the values of $\eta$ calculated 
using the FSS and the CT methods 
for the 2D Heisenberg model. The differences with the 2D XY-model are clear. 
It is particularly noticeable that the 
estimates given by the three 
methods are completely different.
Even the shapes of the curves are different,
and no agreement can be found. In particular, the determinations 
of $\eta$ deduced 
from $m$ and $\chi$, respectively, do not agree with each other.
It follows that  the  hyperscaling relation (\ref{eq:hyper}) 
is definitively not satisfied. 

These observations give a strong evidence 
for the lack of any critical behaviour in
the 2D Heisenberg model within the range of temperatures considered here.

\subsection{The 2D LL-model}
\begin{figure}[!htb]
\centerline{\includegraphics[width=0.45\textwidth]{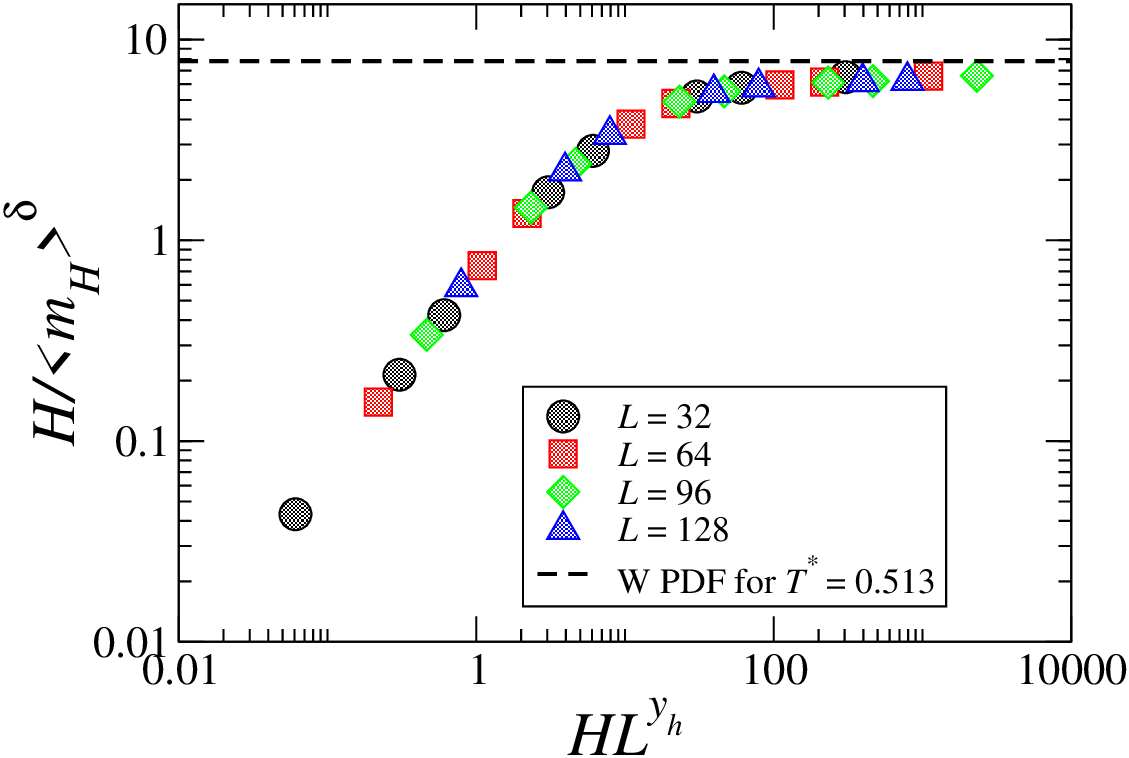}}
\caption{2D Lebwohl-Lasher model in a magnetic field 
($H = 0.1$, $0.01$, $0.05$, $0.001$, $0.005$, $0.0001$, $0.0005$) at $T = 0.513$. 
The bold straight line represents the large $X=HL^{y_h}$ limit for a Weibull-like PDF (W). 
Each point corresponds to an average
over $100,000$ independent realizations.}
\label{fig:fig4ll}
\end{figure}

We present now a similar systematic work, performed on the 2D LL-model.
In particular, the figure  (\ref{fig:fig4ll}) shows results for  the 2D LL-model in an applied  magnetic field
at the expected critical temperature, $T^\star$. 

\begin{figure}[!htb]
\centerline{\includegraphics[width=0.45\textwidth]{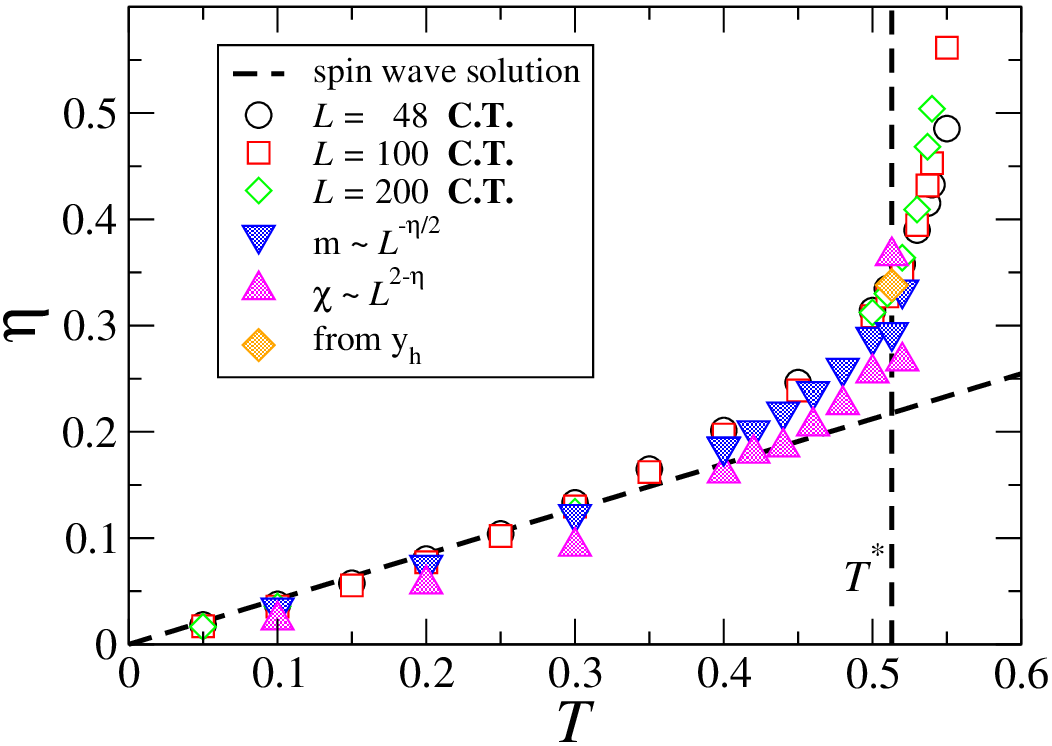}}
\caption{Dependence of the critical exponent $\eta$ with the 
temperature for the 2D Lebwohl-Lasher model. 
(Monte Carlo simulations with $10^6$ MCS/spin after removing the first $10^6$ 
MCS to achieve proper thermalization). 
The conformal transformation (CT) data are from 
reference \cite{FPB}. The dashed line is the spin-wave result 
$\eta=4k_BT/3\pi J$.}
\label{fig:etall}
\end{figure}

The results are consistent with the equation  (\ref{eq:universal}), and with the 
saturation of 
the quantity $Y$. The best collapse was obtained for $\delta = 10.83$. 
Then, using the hyperscaling relation  (\ref{eq:hyper2}) we found: 
$\eta = 0.338$ at this temperature. 
This result is shown in figure  (\ref{fig:etall}) and the agreement 
with the value of $\eta$ obtained from 
the other methods is excellent at this temperature.

In figure  (\ref{fig:etall}), we show the values of $\eta$ for the 2D LL-model 
at various temperatures. 
Estimates were done using FSS of the magnetization 
(from reference \cite{PFB} and figure  \ref{fig:mvslll}) and of 
the magnetic susceptibility (from figure  \ref{fig:chivslll}).
They are compared to the values estimated from the CT method (from reference \cite{FPB}),
and from the collapse of the data in a magnetic field. 

Nevertheless, one should stress that the 
agreement between the values estimated 
by all the four methods is good, but imperfect. 
Small discrepancies can easily be noticed. In particular, the estimates 
which are based on the FSS behaviour of the magnetic susceptibility are 
systematically smaller than those following
from the other methods. The errors are larger than $10\%$, a relevant quantity
probably far out of the range of acceptable error bars. 
In a previous work \cite{RAR},  this error was only of $3\%$, 
but the authors then checked systematically the 
hyperscaling relation, equation  (\ref{eq:hyper}),  
and they concluded that the hyperscaling relation is definitively 
not satisfied at low temperature  (e.g. at $T=0.4)$
(see discussion in the section \ref{sec:PDFhyper} below). 
This has been the first 
evidence for absence of a QLRO phase in the 2D LL-model.

\section{ Revisiting the behaviour of the 2D LL-model at the  `critical' temperature}\label{sec:PDFhyper}

In the previous section, we observed that the critical behaviour of the 2D 
XY-model and 
the behaviour of the 2D LL-model were quite similar. 
Nevertheless, we saw small discrepancies for the LL-model, 
in particular in the behaviour of the second moment of the magnetization, and 
on the validity of the hyperscaling relation (\ref{eq:hyper}). 

In reference \cite{RAR}, the authors performed 
intensive simulations of the 2D LL-model at the temperature $T=0.4$ well 
below the expected critical temperature. 
Clear failure of the hyperscaling relation, failure of collapse of the PDF 
in the first scaling form \cite{bp}, and
analysis of the Binder cumulant, led to the conclusion that 
no QLRO phase exists in the
2D LL-model at this temperature. 

Here, we consider the problem of the existence of a 
critical  at the 
temperature  $T^\star=0.513$, as this temperature
was reported as a possible BKT temperature 
for the 2D LL-model (see section \ref{orderparameter} and 
references \cite{kuzu,FPB}). 
To elaborate on this question, we will report results about the 
probability distribution function, the magnetization scaling 
and the hyperscaling relation (\ref{eq:hyper}) at this temperature. 
We shall also compare the stiffness for the LL- and XY-models.

\subsection{The first-scaling law}

\begin{figure}[!htb]
\centerline{\includegraphics[width=0.45\textwidth]{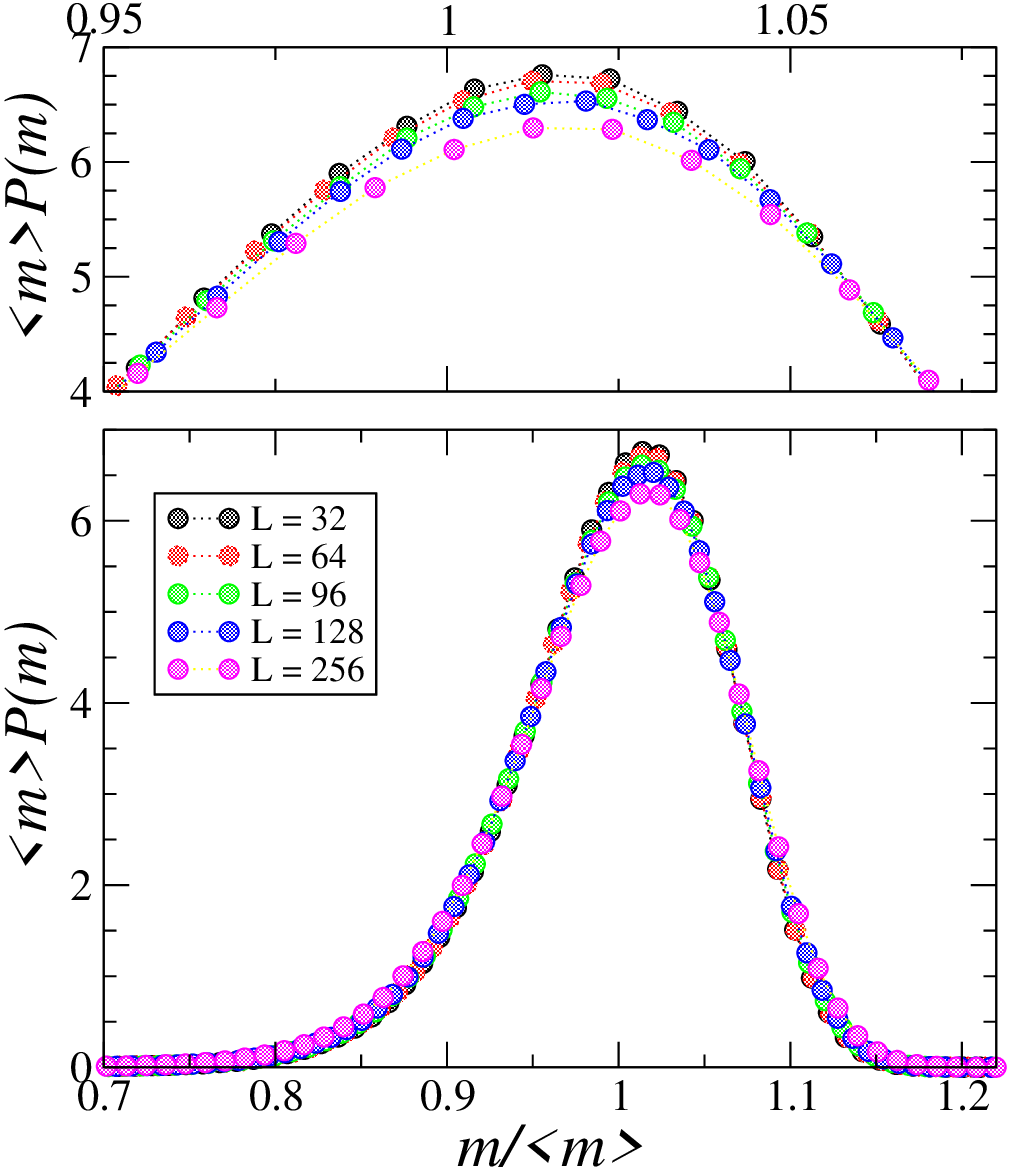}}
\caption{ PDF of the magnetization for the 2D Lebwohl-Lasher model at the
temperature $T^\star=0.513$, plotted in the first-scaling form (\ref{eq:1}). The
scaling law is \emph{not} confirmed for $L=32$ up to $L=256$.
Wolff's single-cluster algorithm was used \cite{wolff}. Each data set
corresponds to average over $6,000,000$ independent realizations.}
\label{fig:pdfll}
\end{figure}

If the 2D Lebwohl-Lasher model is critical at the temperature 
$T^\star = 0.513$, 
self-similarity must be observed at this temperature
because no finite characteristic length can exist at  criticality.
Asymptotic self-similarity results in the so-called first-scaling law for the 
order parameter, here the magnetization $m$, \cite{bp}:
\begin{equation}
\langle m \rangle P(m)\equiv\Phi_{T}(z_1), \;\;\;\;\;\; \mbox{with} 
\;\; z_{1}\equiv\frac{m}{\langle m \rangle}\label{eq:1},
\end{equation} 
with $\Phi_T$ a scaling function which depends only on the actual temperature $T$. 
Equation (\ref{eq:1})  is sequel of the standard finite-size scaling theory \cite{cardy}, 
but it is highly advantageous that equation  (\ref{eq:1}) 
does not require knowledge of the values of any critical exponent. 
Note that, under this form, the hyperscaling relation,
\begin{equation} \label{eq:hypernew}
\frac{\langle m \rangle}{\sigma}\quad=\quad \mbox{ constant term}, 
\end{equation}
is automatically realized with $\sigma$ the standard deviation of the order parameter. 
Indeed, one has: $\sigma^{2} \propto \chi/L^{d}$, and $\sigma$ scales then with the system size 
as: $\sigma \sim L^{\gamma/2\nu-1}$ for  2D systems. Therefore, 
the ratio $\langle m\rangle/\sigma$ should be a constant whenever the hyperscaling relation 
(\ref{eq:hyper}) is satisfied \cite{RAR}.

Figure (\ref{fig:pdfll}) shows the behaviour of the PDF when the system size increases, for the 2D LL-model at 
the temperature $T^\star=0.513$. We can observe failure of the collapse 
between these curves, specially for the largest 
system sizes,
where  the distributions tend to deviate  from each other when $L$ increases. 
One must conclude that the temperature 
$T^\star=0.513$ is \emph{not} critical for this model.

In two others papers published recently, one can find the same kind of 
curve for the 2D XY-model at 
$T = 0.6 < T_{BKT}$ \cite{RAR} and at $T = T_{BKT}$ \cite{rprb}. 
The collapse of all the PDF is excellent in the two cases, 
a result consistent with the known criticality of the system at $T_{BKT}$ 
and all the temperatures below this value.
In the same reference  \cite{RAR}, the 2D LL-model was studied similarly 
at $T = 0.4$.
Failure of the collapse was clear in this case, leading to the conclusion 
that the system is not critical below the assumed transition temperature.

\begin{figure}[!htb]
\centerline{\includegraphics[width=0.45\textwidth]{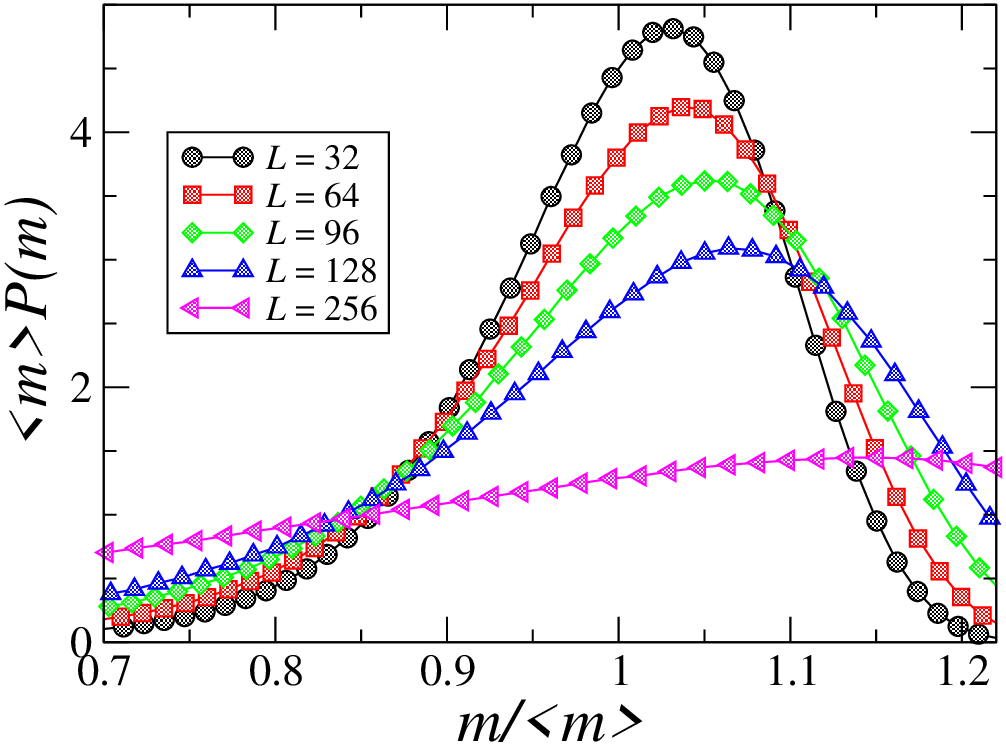}}
\caption{PDF of the magnetization for the Lebwohl-Lasher model at $T=0.548$, plotted in the first-scaling form (\ref{eq:1}). 
Each data set
corresponds to average over $6,000,000$ independent realizations.}
\label{fig:pdfllsobreTc}
\end{figure}

In 2003, Mondal {\it et al.} \cite{mr} reported a possible 
transition temperature for the 2D LL-model at 
the temperature $T = 0.548$, larger than the values reported by 
Kunz {\it et al} in 1992 \cite{kuzu}. 
We thus also
performed numerical simulations of the 2D LL-model at $T = 0.548$. 
In figure  (\ref{fig:pdfllsobreTc}) the data are plotted in the 
first-scaling form (\ref{eq:1}). 
The data are far from collapsing. It is actually similar to the 
behaviour of a system far away from any critical point. 

\begin{figure}[!htb]
\centerline{\includegraphics[width=0.45\textwidth]{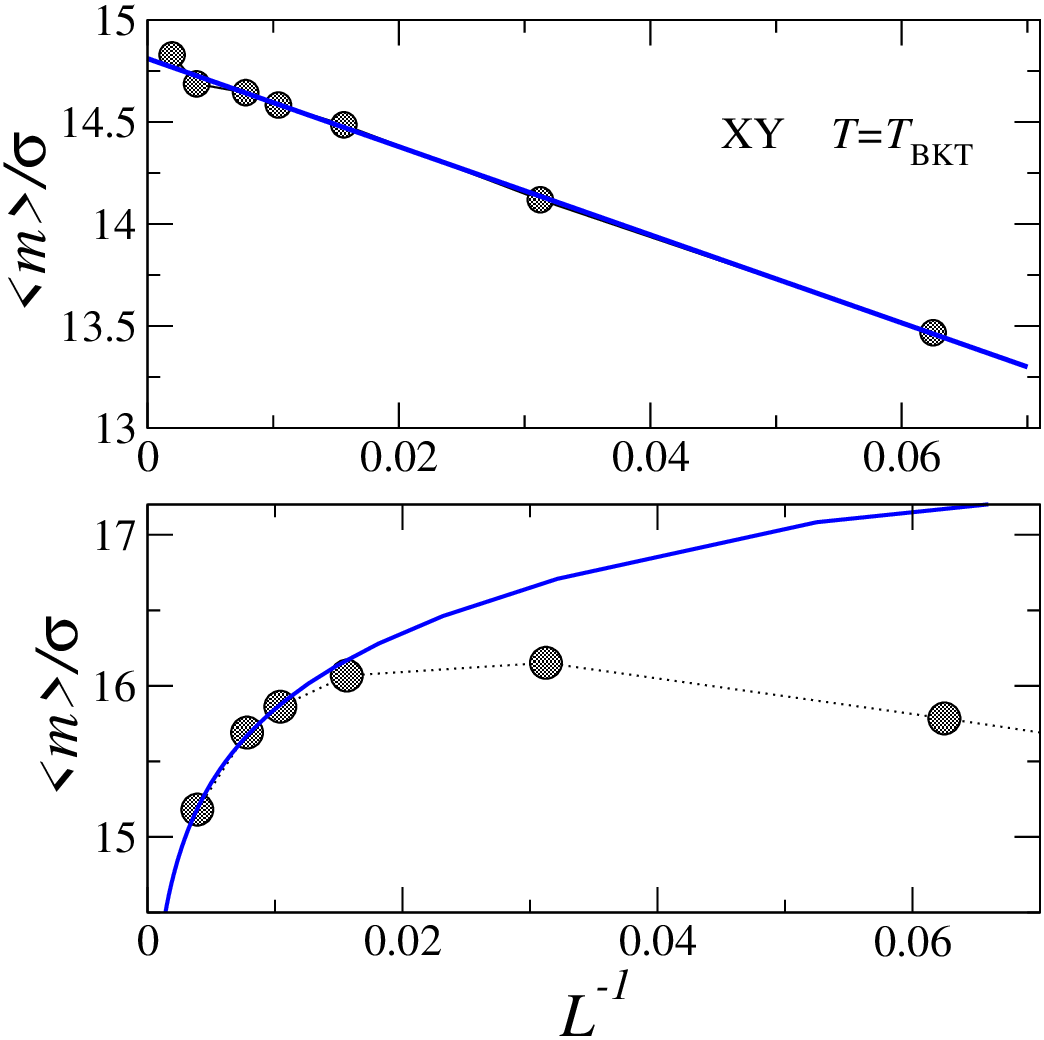}}
\caption{$m/\sigma$ vs. $L^{-1}$ at $T=T_{BKT}$ for the 2D XY-model
(top) and at $T=T^\star=0.513$ for the 2D Lebwohl-Lasher model (bottom). 
For the XY-model, the data were 
obtained from the reference \cite{rprb}. For the LL-model, 
the blue line is a power law fit.}
\label{fig:msigmaxy}\label{fig:msigmall}
\end{figure}

\subsection{The hyperscaling relation}

In this section, we check numerically
the hyperscaling relation under the form (\ref{eq:hypernew}). 
In figure  (\ref{fig:msigmaxy}), 
$\langle m \rangle/\sigma$ is plotted versus $1/L$ for the 
2D XY-model and the 2D LL-model respectively, at the BKT temperatures reported 
or expected for each system, namely: $T_{BKT} = 0.893$ for the XY-model and $T^\star = 0.513$ for the LL-model.

For the XY-model, the ratio is seen to converge to a definite value 
($\simeq 14.8$) at the thermodynamic limit $1/L=0$, 
while for the LL-model, a power law is the best fit consistent with 
the data of $\langle m \rangle/\sigma$ versus $L$,
and this fit leads to a vanishing value of the ratio at the thermodynamic 
limit. 
We conclude that the hyperscaling relation is very likely violated in 
this case, for the obvious reason that 
the system is \emph{not} critical at the temperature  $T^\star=0.513$. 

\subsection{The stiffness modulus}

Appearance of stiffness is an important consequence of the spontaneous symmetry breakdown 
in a system with continuous symmetry. It is closely related to correlations in the 
transverse response function, 
any spatial variation of the order parameter, perpendicular to the direction of the ordering, increasing 
the energy of the system. For this reason, the system generates a restored strength in response 
to any attempt to create a new configuration: this is known as  stiffness.

In the case of the 2D XY-model, this is the helicity 
modulus and it is defined as follows. 
Let
\begin{equation}
{\cal H}=-\beta J \sum_{\langle i,j \rangle } {\bf S}_{i} \cdot \hat {g}_{\alpha}{\bf S}_{j}
\end{equation}
be the modified Hamiltonian in which  $\hat {g}_{\alpha}$ is a rotation of
angle $\phi$
of the system around any axis,
say here the $x$-axis. The difference between the free 
energy $F(\phi)$ calculated with the modified Hamiltonian, 
and the free energy calculated with the original Hamiltonian, 
\begin{equation}
F(\phi)-F(0)=-\frac{1}{\beta}\ln\left(\frac{{\rm Tr\ \!}
{\rm e}^{-\beta {\cal H}_{g_x}}}{
{\rm Tr\ \!}
{\rm e}^{-\beta {\cal H}_{0}}
}
\right),
\end{equation}
varies quadratically with the rotation angle $\phi$, 
at least for $\phi \ll 1$. 
The coefficient, usually written: $\Upsilon$, is the helicity modulus:
\begin{equation}
F(\phi)-F(0)=-\Upsilon\phi^2+O(\phi^4)
\end{equation}
and it takes the form
\begin{equation}
\frac{2L^d}{J\beta} \Upsilon=
\left\langle
\sum_{\langle i,j \rangle } {\bf S}_{i}\cdot \hat J_x^2 {\bf S}_{j} 
\right\rangle
- J\left\langle 
\Bigl( \sum_{\langle i,j \rangle } {\bf S}_{i}\cdot \hat J_x {\bf S}_{j}
\Bigr)^2\right\rangle 
\end{equation}
where $\hat J_x$ is the rotation operator around the $x$-axis.

\begin{figure}[!htb]
\centerline{\includegraphics[width=0.45\textwidth]{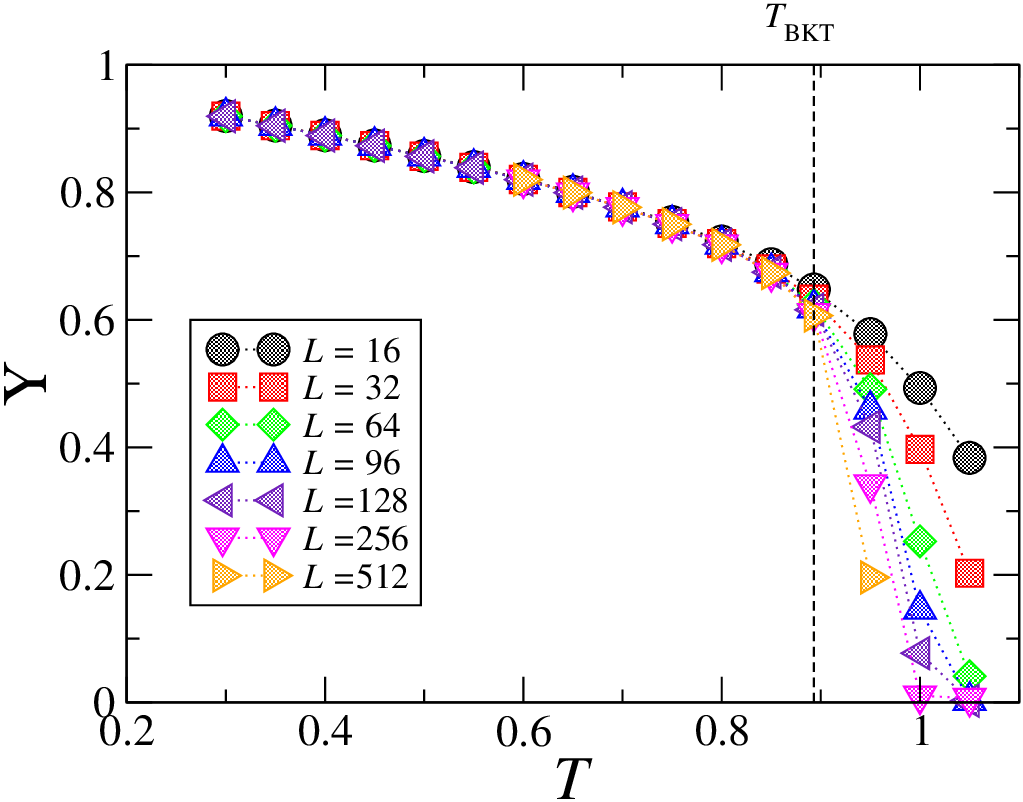}}
\caption{Temperature behaviour of the stiffness modulus $\Upsilon$ for the 2D XY-model.  
(Monte Carlo simulations with $10^6$ MCS/spin after removing the first $10^6$ 
MCS to achieve proper thermalization).}
\label{fig:rigxy}
\end{figure}

\begin{figure}[!htb]
\centerline{\includegraphics[width=0.45\textwidth]{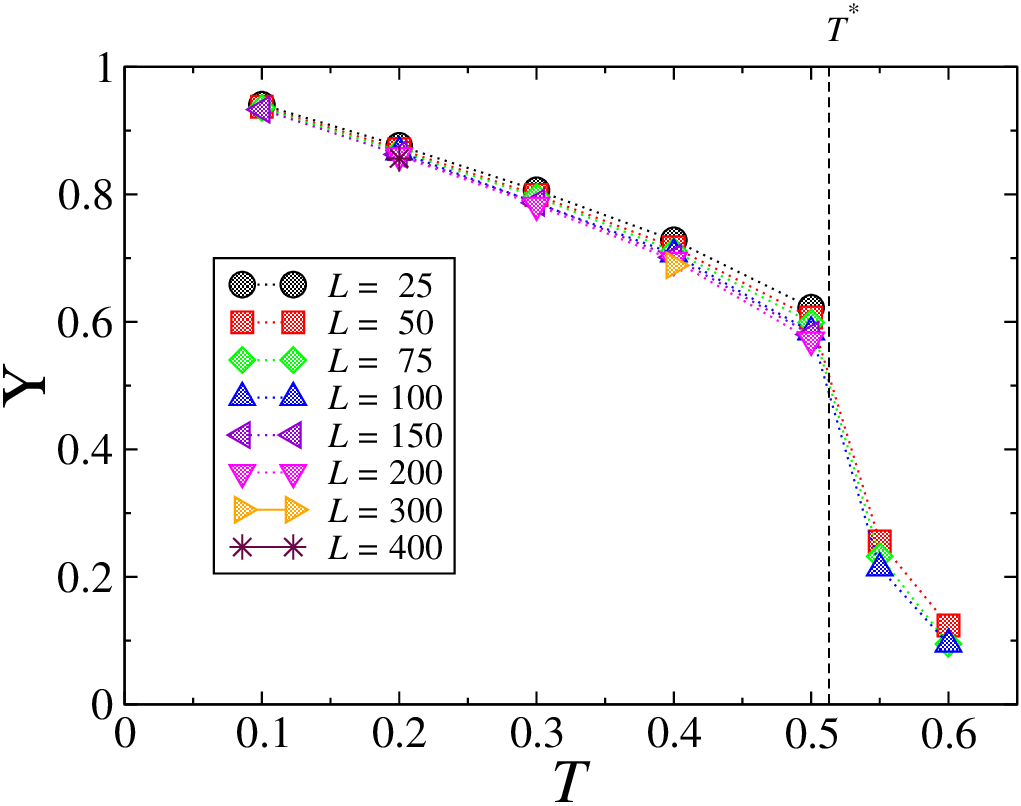}}
\caption{Temperature behaviour of the stiffness modulus $\Upsilon$ for the 2D 
Lebwohl-Lasher model. (Monte Carlo simulations with $10^6$ MCS/spin after removing the first $10^6$ 
MCS to achieve proper thermalization).}
\label{fig:rigll}
\end{figure}

In figure  (\ref{fig:rigxy}) the stiffness modulus for the 2D XY-model is shown versus the temperature
for various sizes from $L=16$ to $L=512$. The behaviour is typical of an order parameter: below $T_{BKT}=0.893$,
the plots are almost independent of 
the system size, and this is the sign of
the rigid phase with a finite value of the stiffness constant, 
while  
for temperatures larger than  $T_{BKT}=0.893$, the system's response
becomes weaker, leading to the decrease of the stiffness modulus.
The high-temperature phase is a disordered phase with the vanishing stiffness.
The change of behaviour in the stiffness modulus between the two phases 
exhibits the appearance 
of a topological phase transition, with quasi-long-range order, 
but finite rigidity, 
in the low temperature phase.

For the 2D LL-model, the same procedure can be used
 with eventually the following expression for the stiffness:
\begin{widetext}
\begin{equation}
\frac{2L^d}{3J\beta} \Upsilon=-\left\langle
\sum_{<i,j>}\left(({\bf S}_{i}\cdot {\bf  S}_{j})\times({\bf  S}_{i}\cdot\hat J_x^2{\bf  S}_{j})
+({\bf S}_{i}\cdot\hat J_x {\bf S}_{j})^2\right)\right\rangle
+3J \left\langle\Bigl(\sum_{<i,j>}
({\bf S}_{i}\cdot {\bf S}_{j})\times({\bf S}_{i}\cdot\hat J_x {\bf S}_{j})\Bigr)^2\right\rangle 
\label{eq:mtcsigma}
\end{equation}
\end{widetext}

\begin{figure}[!htb]
\centerline{\includegraphics[width=0.45\textwidth]{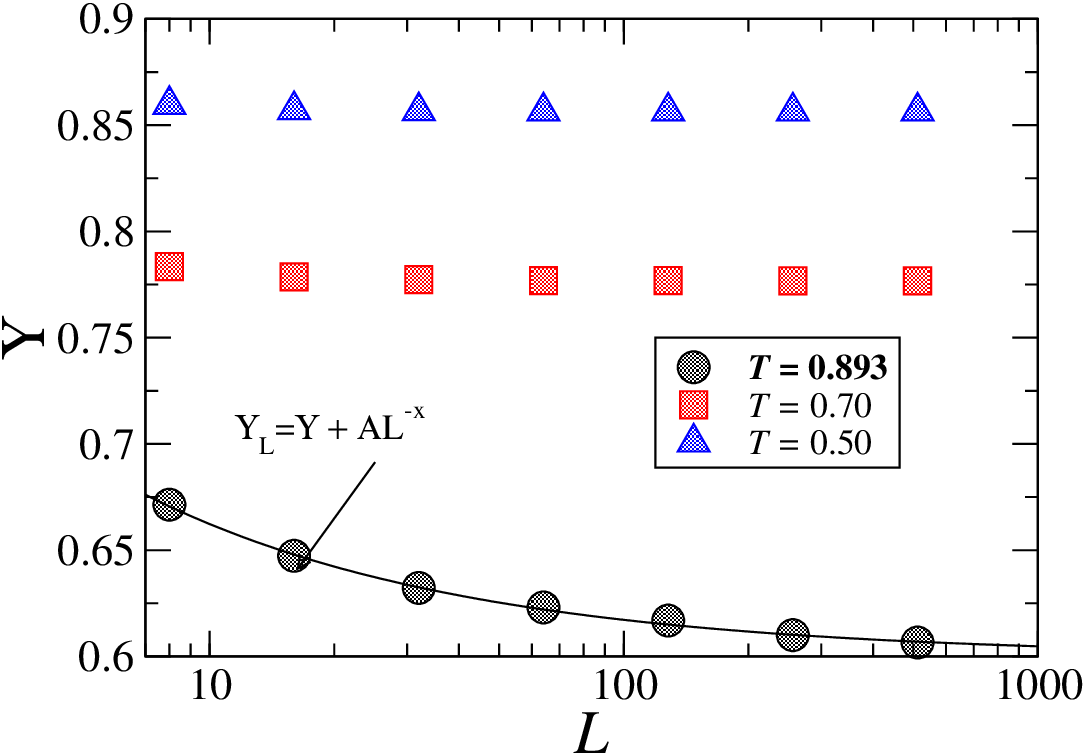}}
\caption{ Stiffness $\Upsilon$ versus system size $L$ for three different temperatures for the 2D XY-model. $\Upsilon(L)$ goes to a finite value in the limit $L\rightarrow \infty$ for $T\le T_{BKT}$.  (Monte Carlo simulations with $10^6$ MCS/spin after cancellation of $10^6$ for thermalization).}
\label{fig:rigvsLXY}
\end{figure}

Then, we have plotted $\Upsilon$ versus the temperature 
as it is shown in figure  (\ref{fig:rigll}) 
for various sizes from $L=25$ to $L=400$. 
Again, a behaviour similar to an order-parameter  
is observed in this case. 
At any temperature below  $T^\star=0.513$, one clearly identifies finite values of the stiffness
independently of the system size. Above $T^\star$, the values of the stiffness drop to values close to $0$.

However, beyond the similarities, one can remark that the stabilization to a finite value of the
stiffness in the low-temperature phase, is not similar in both models. 
One can see this point, in particular analyzing 
the dependences with the system size. 
In figure  (\ref{fig:rigvsLXY}), $\Upsilon$ is shown versus $L$ for three 
different temperatures of the 2D XY-model. 
Two of them below the BKT transition ($T=0.5$ and $0.7$), and the other 
temperature just at the BKT transition. 
For the lower temperatures $\Upsilon$ goes fast to an asymptotic constant 
value. 
At $T_{{BKT}}$, the stiffness modulus reaches its saturation value 
algebraically. 
Then, it is clear that for the XY-model a sharp jump of the stiffness 
modulus is observed right at the topological transition. 
\begin{figure}[!htb]
\centerline{\includegraphics[width=0.45\textwidth]{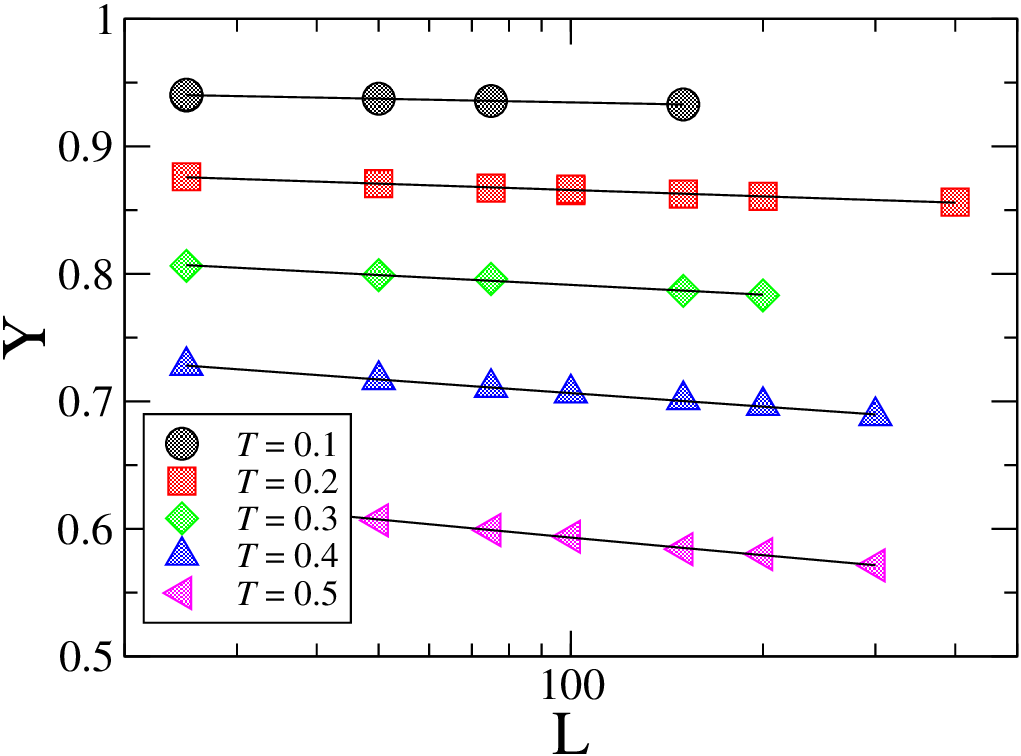}}
\caption{Behaviour of stiffness modulus versus the system size for the 2D Lebwohl-Lasher model. 
(Monte Carlo simulations with $10^6$ MCS/spin after removing the first $10^6$ 
MCS to achieve proper thermalization).}
\label{fig:rigvsLLL}
\end{figure}
The behaviour is completely different for the 2D LL-model.
 In figure  (\ref{fig:rigvsLLL}), the dependence 
of $\Upsilon$ with the system size $L$ is shown for various temperatures 
below $T^\star$.
The stiffness does not  
reach a finite value in the limit $L\rightarrow \infty$. 
This behaviour is similar to the 2D Heisenberg 
model \cite{MouhannaEPL} and to the fully frustrated Heisenberg 
model \cite{Wintel}.
In figure  (\ref{fig:rigvsLLL}), the slopes of the curves depend on 
the value of the temperature. 
It appears to be similar to the results of Winttel et al \cite{Wintel}. 
A detailed study of the stiffness for frustrated spin systems can be 
found in \cite{MouhannaPRB}.

\section{Discussion and Conclusions}
Our initial goal was to give definite conclusions about the possible appearance of a critical topological transition,
for example of the Berezinskii-Kosterlitz-Thouless type, in the 2D Lebwhol-Lasher model.

Our results exhibit a scenario more complex than expected. It can be summarized as follows:
\begin{itemize}
\item  
The behaviour of the order parameter (the magnetization) with the temperature and the system size,
pointed out a quasi-long range order phase for temperatures below the value $T^\star = 0.513$. 
One has noticed an excellent agreement of the correlation exponent $\eta$ estimated using conformal transformation, 
scaling of the order parameter with $L$, including scaling with a magnetic field. 
\item
If we investigate a quantity related to the second moment of the order parameter, 
that is the susceptibility, a  
\emph{qualitative} agreement  with a QLRO phase behaviour is observed.
However, the value obtained for $\eta$ from this 
quantity does not compare well with 
the estimates from the order parameter. Actually, the hyperscaling 
relation (\ref{eq:hyper}) 
is not satisfied for $T\le T^\star$. This is in clear
contradiction with the possibility of 
a line of critical point below $T^\star$. 
Then, one has to conclude that a BKT transition is not observed for this model.
\item
Although not studied in the present paper, 
the behaviour of the Binder cumulant (which depends on the fourth 
moment of the order parameter), as 
reported in \cite{RAR}, fully supports  the above conclusion: 
no universality is observed at any finite temperature. 
Signs of QLRO just disappear if larger moments of the order parameters are used. 
\item
As the conclusion seems to depend on the order of the moment of 
the magnetization, 
it looks natural to use the complete probability distribution 
function of the order parameter
to clarify the situation. Unlike the 2D XY-model, the 2D LL-model 
does not show any first-scaling-law collapse 
as it would be expected 
for a critical system. The absence of self-similarity 
at any temperature discards any sort of transition in this model, which
thus appears more similar to the Heisenberg model than to the XY-model
eventually.
\item The stiffness modulus mixes a second and a fourth moment of the 
order parameter. 
For the 2D LL-model, this quantity does not exhibit an {\em order-parameter 
behaviour}. 
At  low temperatures a logarithmic decay with $L$ is observed, 
and asymptotic finite values
are highly unlikely, resulting in a non rigid phase at low temperatures.  
\end{itemize}

Our conclusion, based only on numerical simulations, 
is that the 2D LL-model shares similarities with the fully frustrated
 antiferromagnetic Heisenberg model 
(see \cite{Wintel,Kawamura}). Then, the overall behaviour could well be 
a sharp crossover \cite{Wintel}, instead of a real  
transition to a critical phase, 
a new kind of frustration preventing the topological defects 
of the LL-model to initiate a true phase transition.

\section*{ Acknowledgment} We gratefully acknowledge Yu. Holovatch for
discussions.

\label{last@page}

\begin{thebibliography}{2000}
\bibitem{wolffH} For the 2D XY-model or 2D LL-model in the magnetic field $\bf H$, we  used a Wolff algorithm similar to the one used for the ${\bf H}=0$ case, with an additional spin which interacts with all the other spins, with $\bf H$ as the strength of the interaction.
\end{thebibliography}
\end{document}